\documentclass[aps,prb,twocolumn,preprintnumbers,amsmath,amssymb,superscriptaddress]{revtex4-1}

\usepackage{amsmath,amssymb,amsfonts,mathrsfs} 
\usepackage{amsthm}
\usepackage{CJK}
\usepackage{bm}
\usepackage{graphicx}
\usepackage{color}
\usepackage[normalem]{ulem} %strikeout word \sout

\usepackage{hyperref,cleveref}

\def \beq {\begin{equation}}
	\def \eeq {\end{equation}}

\def \K {\hat{\mathcal{K}}}
\def \Z {\mathbb{Z}}
\def \H {\mathcal{H}}
\def \A {\mathcal{A}}

\def \T {\hat{\mathcal{T}}}

\def \P {\hat{\mathcal{P}}}

\def \y {\hat{\mathbf{y}}}
\def \z {\hat{\mathbf{z}}}
\def \k {\mathbf{k}}
\def \e {\hat{\mathbf{e}}}

\begin{document}

\title{Topological transport in Dirac nodal-line semimetals }

\author{W. B. Rui}
\affiliation{Max-Planck-Institute for Solid State Research, Heisenbergstrasse 1, D-70569 Stuttgart, Germany}

\author{Y. X. Zhao}
\email[]{zhaoyx@nju.edu.cn}
\affiliation{National Laboratory of Solid State Microstructures and Department of Physics, Nanjing University, Nanjing 210093, China}
\affiliation{Collaborative Innovation Center of Advanced Microstructures, Nanjing University, Nanjing 210093, China}

\author{Andreas P.\ Schnyder}
\email[]{a.schnyder@fkf.mpg.de}
\affiliation{Max-Planck-Institute for Solid State Research, Heisenbergstrasse 1, D-70569 Stuttgart, Germany}

\date{\today}

\begin{abstract}
Topological nodal-line semimetals are characterized by one-dimensional Dirac nodal rings that are protected
by the combined symmetry of inversion  $\mathcal{P}$  and time-reversal  $\mathcal{T}$.
The stability of these Dirac rings is guaranteed by a quantized $\pm \pi$ Berry phase and their
 low-energy physics is described by a one-parameter family of (2+1)-dimensional quantum field theories
 exhibiting the parity anomaly. Here we study the Berry-phase supported topological transport of  $\mathcal{P}\mathcal{T}$ invariant nodal-line semimetals.
 We find that small inversion breaking allows for an electric-field induced anomalous transverse current, whose universal component originates
 from the parity anomaly. Due to this Hall-like current, carriers at opposite sides of the Dirac nodal ring flow to opposite surfaces when an electric field is applied.
To detect the topological currents,  we propose a dumbbell device, which uses surface states
to filter charges based on their  momenta. Suggestions for experiments and  device applications are discussed.
\end{abstract}

\vspace{-0.4cm}


\maketitle

The last decade witnessed a growing interest in anomalous transport properties of topological semimetals~\cite{Classification-RMP,Hosur_Weyl_develop,book_zee_quantum_field,ryu_moore_ludwig_PRB_12,burkov_review_IOP,vishwanath_PRX_14}, such as  
the axial current in Weyl semimetals~\cite{zyuzin_burkov_PRB_12} and the valley Hall effect in graphene~\cite{XiaoYao-Valley-Hall,Niu-RMP}. These topological currents have their origin in quantum anomalies of the relativistic field theories describing the low-energy physics of semimetals.
Quantum anomalies arise whenever a symmetry of the classical theory is broken
by the regularization of the quantum theory. 
For example, in Weyl semimetals the (3+1)-dimensional chiral anomaly~\cite{ChiralAnomaly-A,ChiralAnomaly-BJ,nielsen_ABJ_anomaly_weyl,vivek_ABJ_PRB_12,son_spivak_PRB_13,yang_lu_ying_PRB_11,liu_qi_PRB_13} manifests itself by the non-conservation
of the chiral charge, i.e., as an axial current flowing between Weyl points with opposite chiralities.  Experiments on TaAs~\cite{Chiral-Anomaly-Exp-I,Chiral-Anomaly-Hasan}
and on Na$_3$Bi~\cite{Xiong413} have revealed signatures of the chiral anomaly in magneto-transport measurements. 
The chiral anomaly of Weyl semimetals is intimately connected to 
the nontrivial topology of the Berry bundle~\cite{Horava-FS,Volovik-book,volovik-Vacuum,ZhaoWang-Classification}, which endows the Weyl points with a nonzero topological charge.

Another example of an anomaly leading to topological currents is the (2+1)-dimensional parity anomaly~\cite{Semenoff-ParityAnomaly,Redlich-ParityAnomaly,Haldane1988,dunne_arXiv}, 
which is realized in graphene~\cite{Graphene-superlattices,Bilayer-Graphene-I,Bilayer-Graphene-II,Valley-Hall-MoS2-I,Valley-polarization-MoS2-I,Valley-polarization-MoS2-II}.  
The parity anomaly   also appears on the surface of topological (crystalline) insulators~\cite{Fu_Kane,fang_rotation_2017} and in quantum spin Hall systems~\cite{wieder_wallpaper_2017}. 
The fermionic excitations near the Dirac cones of graphene 
are described by a (2+1)-dimensional  quantum field theory exhibiting the parity anomaly. 
Any gauge symmetric regularization of this quantum field theory must break spacetime inversion symmetry,
which manifests itself by a parity-breaking Chern-Simons term  in the electromagnetic response theory of a single graphene Dirac cone.
This Chern-Simons term gives rise to the   valley Hall effect, where fermions from different Dirac cones flow to opposite transverse edges,
 upon applying an electric field. The valley Hall effect has been observed experimentally~\cite{Graphene-superlattices,Bilayer-Graphene-II,Bilayer-Graphene-I,Valley-Hall-MoS2-I}
 and has attracted attention due to possible applications in valleytronics devices~\cite{Graphene-superlattices,Valley-Filter_2007}.
 
 Parallel to these developments, recent research has shown that there exist topological semi-metals
 not just with Fermi points, but also with \emph{finite-dimensional} Fermi surfaces, such 
 as, Dirac or Weyl nodal lines~\cite{Volovik-book,volovik-Vacuum,Horava-FS,ZhaoWang-Classification,burkov_balents_PRB,matsuura_NJP_13,Sato-K1,Reflection-SM,Nodal-Line-I,Nodal-Line-II,Yamakage,zhao_schnyder_PRB_16,Zhao-Schnyder-Wang-PT,fang_topological_2015,Wieder_and_Kane,li_dirac_2017}.  These line nodes can be protected by time-reversal and nonsymmorphic symmetries~\cite{fang_topological_2015,Wieder_and_Kane}, by mirror planes~\cite{Wieder_and_Kane}, or by the combined symmetry of inversion and time-reversal~\cite{Zhao-Schnyder-Wang-PT,fang_topological_2015}.
 The topological charges of these finite-dimensional   Fermi surfaces are defined in a similar way as for Weyl and Dirac points, namely,  by 
 the topology of the Berry bundle on a $d_c$-dimensional sphere  that encloses  the Fermi surface from its transverse dimension~\cite{Volovik-book,volovik-Vacuum,Horava-FS,ZhaoWang-Classification}. Here, $d_c$ is called the co-dimension of the topological Fermi surface. 
Since topologically nontrivial Berry bundles are closely connected to quantum anomalies, one may wonder whether the quantum field theories describing nodal-line semimetals exhibit any anomalies and, if so, whether they lead to unusual transport phenomena.

This is the question we address in this Rapid Communication for the case of Dirac nodal-line semimetals (DNLSMs) 
protected by the combined symmetry of time-reversal  $\mathcal{T}$  and inversion  $\mathcal{P}$  with \mbox{$(PT)^2=+1$}~\cite{Nodal-Line-I,Nodal-Line-II,Yamakage}. $\mathcal{P}\mathcal{T}$ invariant DNLSMs are realized in several different materials, e.g., in Ca$_3$P$_2$~\cite{xie_schoop_Ca3P2_apl_15} and
 CaAgAs~\cite{okamoto_takenaka_JPSJ_16,emmanouilidou_PRB_17}   %with weak spin-orbit coupling,
and in other systems~\cite{NLWeng,NLGraphite,DWZ-ColdAtom-PT,wu_dirac,YuRui-DNLSM,RuiYu-PT-DNLSM,mikitik_PRL_04,hirayama_topological_2017}.
We find that the low-energy fermionic excitations of these DNLSMs are described by a one-parameter family of (2+1)-dimensional quantum field theories with a
parity anomaly. We show that this parity anomaly leads to a Hall-like topological 
current, which can be controlled using electric fields. To detect this anomalous current, we propose a dumbbell-shaped device,
which utilizes the drumhead surface states of DNLSMs to filter electrons based on their momenta.

\begin{figure}
\includegraphics[scale=0.32]{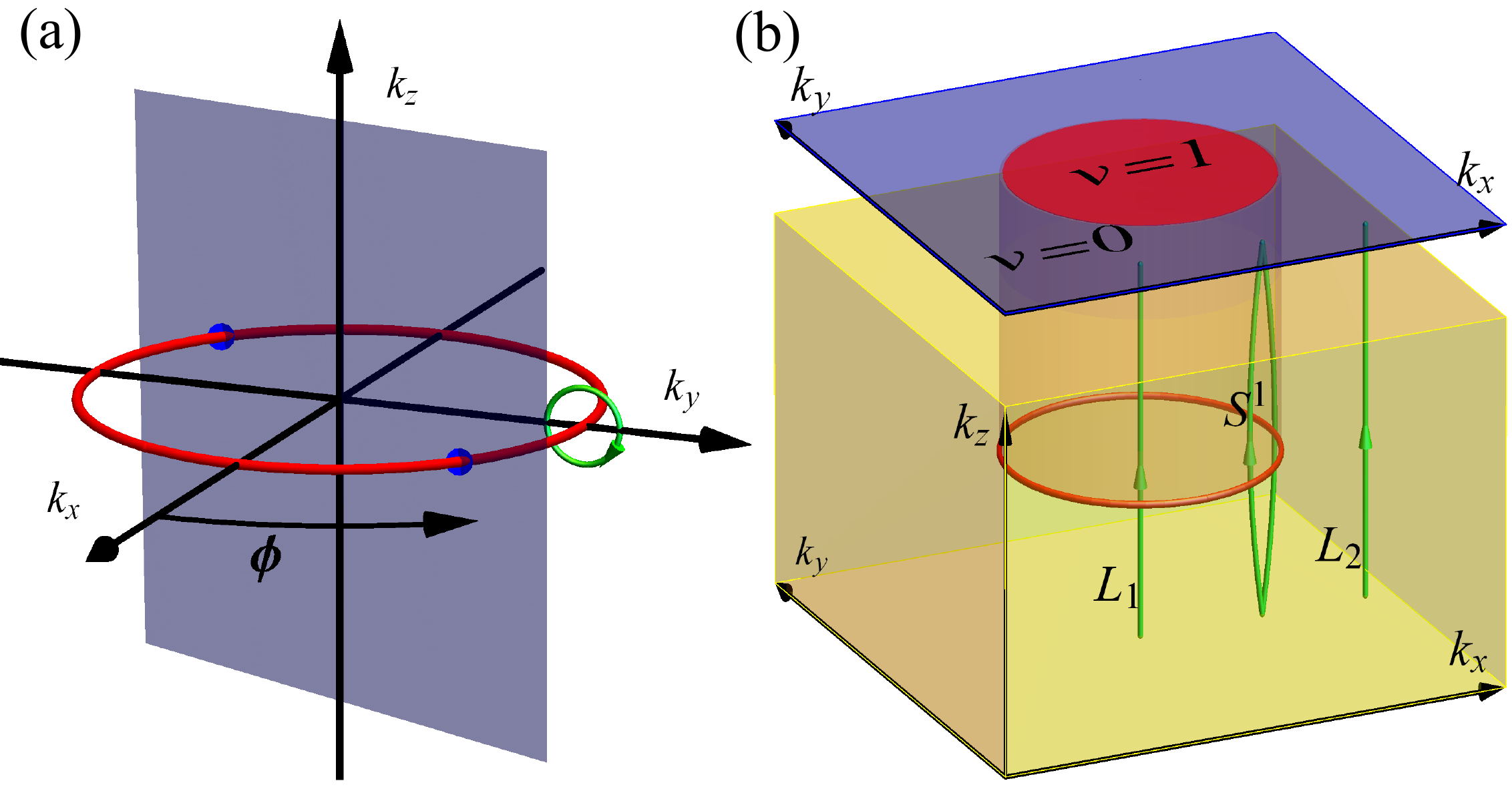} 
\caption{
 %Dirac ring and drumhead surface states.
(a) The topological charge is defined in terms of a line integral along the green loop. The blue plane indicates the two-dimensional subsystems that are parametrized by the angle $\phi$. (b) 
Relationship of the Dirac ring to the surface states of  a topological nodal-line semimetal.
Drumhead surface states occur within the red region, which is bounded by the projected Dirac ring. Within this region
the topological charge $\nu$, Eq.~\eqref{Berry-Phase}, takes on the value $\nu=1$.
\label{Nodalloop}}
\end{figure}

\textit{Topological charge and parity anomaly. } 
 We begin our analysis by discussing the relation between the $\Z_2$ topological charge of $\mathcal{P}\mathcal{T}$ symmetric DNLSMs and the parity anomaly. 
The Fermi surface of Dirac nodal-line semimetals consists of   one-dimensional Dirac rings, which have co-dimension $d_c=1$ in the three-dimensional 
Brillouin zone (BZ).
We assume that the DNLSM exhibits only a single Dirac ring, which is located within  the $k_z = 0$ plane [Fig.~\ref{Nodalloop}(a)].
Its low-energy Hamiltonian reads~\cite{Nodal-Line-II}
\begin{equation} \label{model_ham}
\H( {\bf k} )=\frac{1}{\Lambda}[k_0^2-(k_x^2+k_y^2)-b^2 k_z^2]\sigma_3+ v_z k_z\sigma_2+ m \sigma_1 ,
\end{equation}
where for later use we have introduced a small $\mathcal{P}\mathcal{T}$ breaking
mass $m \sigma_1$. In a DNLSM material this mass term could be generated, for example, by inversion breaking uniaxial strain, pressure  or an external electric field. 
In the absence of $m\sigma_1$ the Hamiltonian $\H( {\bf k} )$ is $\mathcal{P}\mathcal{T}$ symmetric with the 
$\mathcal{P}\mathcal{T}$ operator $\P\T=\sigma_3\K$.
The symmetry protection of the Dirac  ring~\eqref{model_ham} is guaranteed by a quantized $\Z_2$ topological charge $\nu$, which is given by the parity of the Berry phase along
a loop $S^1$ that interlinks with the Dirac ring [Fig.~\ref{Nodalloop}(a)]. That is, $\nu$ is expressed as 
\begin{equation}
\nu [ S^1] =\frac{1}{\pi  }\int_{S^1} d\varphi~\mathrm{tr}\A(\varphi)\mod 2, \label{Berry-Phase}
\end{equation}
where the integration is along the loop $S^1$, parametrized by $\varphi\in[-\pi,\pi)$, 
and $\A_{\alpha\beta,j}=\langle \alpha, \k| i  \partial_{k_j}|\beta,\k\rangle$ denotes the Berry 
connection of the occupied Bloch eigenstates $|\alpha,\k\rangle$.
$\mathcal{P}\mathcal{T}$ symmetry ensures that $\nu$ can only take on the quantized values $0$ and $1$. 
Loops $S^1$ that interlink with a Dirac ring have a nontrivial Berry bundle, which results in
a nonzero topological charge  $\nu=1$.
In two dimensions, Eq.~\eqref{Berry-Phase} assures the stability of the Dirac points in graphene.
In fact, since graphene is $\mathcal{P}\mathcal{T}$ symmetric and its Dirac points have co-dimension $d_c=1$,
 it belongs to the same entry~\cite{SymClass} in the classification of topological semimetals as DNLSMs~\cite{Zhao-Schnyder-Wang-PT}. 
  
 Guided by this observation, we introduce   cylindrical coordinates $\{ k_\rho, \phi, k_z \}$ and decompose the (3+1)-dimensional DNLSM into a family of  (2+1)-dimensional subsystems parameterized 
 by  $\phi$, as shown in Fig.~\ref{Nodalloop}(a). 
  Each subsystem exhibits two Dirac points with opposite Berry phase \footnote{To see this, one may move the green integration loop in Fig.~\ref{Nodalloop} along the Dirac ring from one Dirac point to the other. This demonstrates that the green loop encloses the two Dirac points with opposite orientations.}.
The low-energy physics of a single Dirac point in a given subsystem
  is described by a (2+1)-dimensional quantum field theory with the  action
 \begin{eqnarray} \label{eq_action}
S^{\phi} 
=
\int d^3 x  \, \bar{\psi} 
\left[
i \gamma^\mu ( \partial_\mu + i e A_\mu ) + m 
\right] \psi, 
\end{eqnarray}
where $\psi$ is a two-component Dirac spinor  coupled to the electromagnetic gauge field $A_\mu$. 
Here, $\bar{\psi} = \psi^{\dag} \gamma^0$, $\left\{ \gamma^{\mu}, \gamma^{\nu} \right\} = 2 \eta^{\mu \nu}$, and 
$\eta^{\mu \nu} = \mathrm{diag} ( 1, -1, -1)$.
The mass term $m \bar{\psi} \psi$ breaks spacetime inversion symmetry, since the spinors transform 
under $\mathcal{P}\mathcal{T}$ as $\psi \to \gamma^2 \gamma^0 \psi $  
and  $\psi^{\dag}  \to  - \psi^{\dag} \gamma^0 \gamma^2$.
In the absence of the mass term $m \bar{\psi} \psi$, Eq.~\eqref{eq_action} is $\mathcal{P}\mathcal{T}$ symmetric  
and can be viewed as a classical action of 
$(2+1)$-dimensional Dirac fields. 
It is however impossible to quantize this classical action without breaking the spacetime inversion symmetry, i.e.,  $\mathcal{P}\mathcal{T}$ symmetry is broken by the regularization of the quantum theory. To see this, let us consider 
the Pauli-Villars regularization of the 
effective action $S^{\phi}_{\textrm{eff}} [A, m]$ of Eq.~\eqref{eq_action},  
which is obtained from the fermion determinant by integrating out the Dirac spinors. 
The effective action with zero mass $S^{\phi}_{\textrm{eff}} [A,   0]$ needs to be regularized due to ultraviolet divergences,
which can be achieved by the standard Pauli-Villars method, i.e., 
$S^{\phi, \textrm{reg}}_{\textrm{eff}} [A ] 
= S^{\phi}_{\textrm{eff}} [A,   0] - \lim\limits_{M \to \infty} S^{\phi}_{\textrm{eff}} [A, M]$.
While this regularization scheme preserves gauge symmetry, it breaks $\mathcal{P}\mathcal{T}$ invariance, since the Pauli-Villars mass   
$M \bar{\psi} \psi$  leads in the $M \to \infty$ limit to   the Chern-Simons term~\cite{Redlich-ParityAnomaly,dunne_arXiv} 
 \begin{eqnarray} \label{CS_term}
 S^{\phi}_{\textrm{CS}} =  \eta   \frac{e^2}{4 \pi} \int d^3 x \, \epsilon^{\mu \nu \lambda} A_{\mu}\partial_\nu A_{\lambda},
 \end{eqnarray}
 where $\eta = \pm 1$ is the sign of the Dirac point Berry phase. As discussed in Eq.~\eqref{Berry-Phase}, the Berry phase 
 $\eta$ is equal to the topological charge $\nu$ (up to a sign convention). 
 %is related to the topological charge $\nu$ via $\nu=\eta\mod 2$. 

 %%%%%%%%%%%%%%%%%%%%%%%%%%%%
%%%%%%%%%%%%%%%%%%%%%%%%%%%%

From the modern condensed matter viewpoint, the parity anomaly is attributed to the  $\Z_2$ topological charge $\nu$ of the $\mathcal{P}\mathcal{T}$ symmetric Dirac point. That is, because of the topological obstruction from the nontrivial topological charge, there exists no $\mathcal{P}\mathcal{T}$ symmetric lattice  ultraviolet regularization for a single (2+1)-dimensional  Dirac point. I.e.,  any lattice regularization has to involve an \emph{even} number of nontrivial Dirac points, since the  sum
over all topological charges in the BZ torus must be zero.
This is consistent with the $\Z_2$ nature of the parity anomaly, since   a doublet of  (2+1)-dimensional Dirac points  coupled to gauge fields can be quantized without breaking  $\mathcal{P}\mathcal{T}$ symmetry. 

To conclude, in the process of quantizing the classical action~\eqref{eq_action} we have broken $\mathcal{P}\mathcal{T}$ symmetry due to the
Chern-Simons term~\eqref{CS_term}. 
Thus, although the parity anomaly  strictly speaking occurs only in (2+1) dimensions, it also appears
in (3+1)-dimensional DNLSMs.

\textit{Topological transport in DNLSMs. }
Next we discuss the anomalous transport phenomena that are associated with the parity anomaly. 
Varying the Chern-Simons term~\eqref{CS_term} with respect to the electromagnetic gauge field $A_{\mu}$ yields
the anomalous transverse current  
\begin{equation}  \label{anomalous-current}
j^{\mu}_{\textrm{t}, \phi}  = \eta \frac{e^2}{4\pi}\epsilon^{\mu\nu\lambda}\partial_\nu A_\lambda 
\end{equation}
for a single Dirac cone in a given (2+1)-dimensional subsystem. 
Thus, electromagnetic fields projected onto a two-dimensional subsystem induce a topological current, which flows perpendicular to the applied
field.  
Since the energy bands of   DNLSMs  are, to a first approximation, nondispersive along
the $\phi$ direction,  
one might expect that the electromagnetic response of DNLSMs in the presence of a small $\mathcal{P}\mathcal{T}$ breaking term is dominated
by this topological current.
However, for each two-dimensional subsystem there are two Dirac points that contribute to the transverse current
with opposite signs $\eta = \pm 1$.  Since these two contributions cancel out to zero, the topological current
can only be measured by a device that filters electrons based on their momenta.
 
%%%%%%%%%%%%%%%%%%%%%%%%%%%%
\begin{figure}
\includegraphics[scale=0.62]{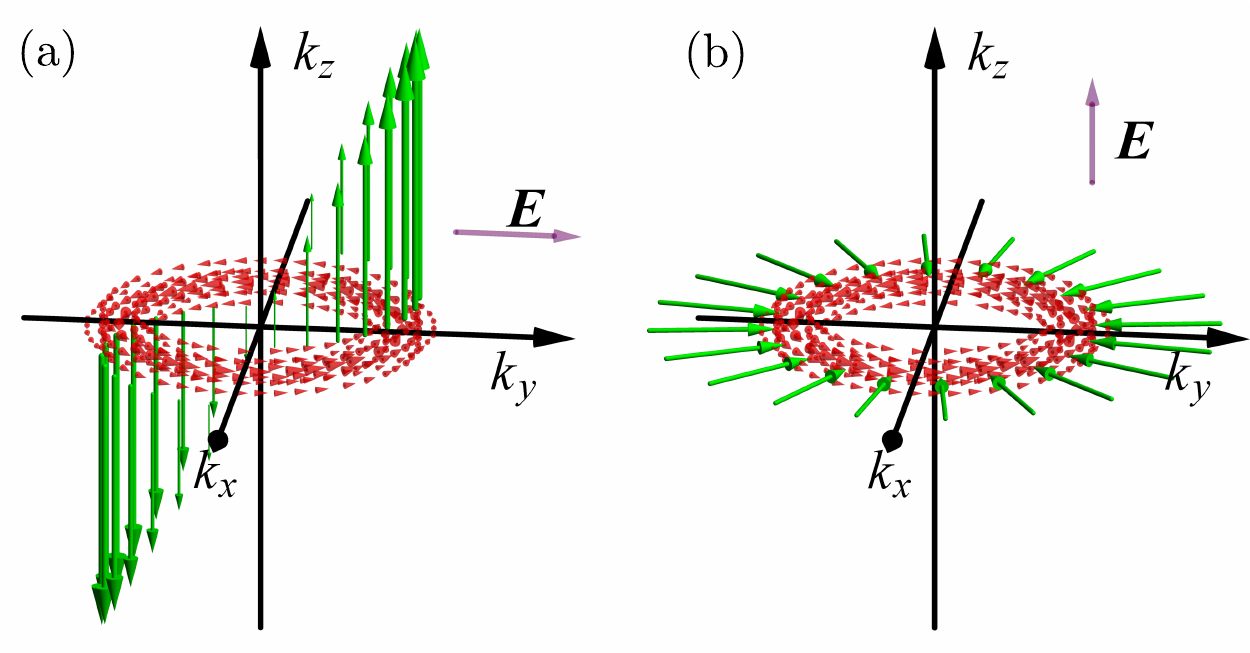} 
\caption{
The red arrows indicate the Berry curvature $\mathbf{\Omega} ({\bf k}) $, Eq.~\eqref{Berry-curvature}, in the presence
of a small $\mathcal{P}\mathcal{T}$ breaking mass term $m \sigma_1$.
The green arrows represent the transverse topological current $  {\bf j}_{\textrm{t}, \phi}$, Eq.~\eqref{Current-density}, that is induced by an external electric field applied along (a) the  $\y$ direction and (b) the $\z$ direction.   \label{Current}}
\end{figure}
%%%%%%%%%%%%%%%%%%%%%%%%%%%%

Let us now give a second derivation of the  transverse topological currents in terms of semiclassical response theory~\cite{Niu-RMP}.
In the presence of an electric field, the semiclassical equations of motion for Bloch electrons contain an anomalous velocity proportional
to the Berry curvature. 
This gives rise to a transverse Hall-like current~\cite{XiaoYao-Valley-Hall,Niu-RMP}, given by
 $
\mathbf{j}_{\textrm{t}}=\frac{e^2}{\hbar}\int \frac{d^3k}{(2\pi)^3} f(\k) ~\mathbf{E}\times \mathbf{\Omega}  (\k)
$,
where $f(\k)$ is the Fermi-Dirac distribution function, $\mathbf{E}$ denotes the electric field, 
and $\mathbf{\Omega} ({\bf k}) $ represents the Berry curvature of the  Bloch eigenstate $|\alpha,\k\rangle$, 
which is defined as $\mathbf{\Omega} ({\bf k})  = \mathbf{\nabla}_{\bf k} \times \langle \alpha, \k | i  \mathbf{\nabla}_{\bf k} | \alpha,\k\rangle $.
From a symmetry analysis it follows that the Berry curvature in a gapped system vanishes identically, unless either time-reversal or inversion symmetry are broken.
Indeed, using Eq.~\eqref{model_ham} with $m=0$ we find that $\mathbf{\Omega} ({\bf k})$ is zero in the entire BZ, except at the Dirac nodal line, where it becomes
singular, i.e., $\mathbf{\Omega} ({\bf k} )=\pi \delta(k_\rho- k_0) \delta(k_z)\e_\phi$.
To regularize this divergent Berry curvature, $\mathcal{P}\mathcal{T}$ symmetry needs to be broken, for example, by uniaxial strain, pressure, disorder, circularly polarized light, or an electric field, 
which leads to a small non-zero mass  $m \sigma_1$ in Eq.~\eqref{model_ham} and, consequently, a well-behaved Berry curvature.
For the conduction band, $\mathbf{\Omega} ({\bf k} )$ is given by  
\begin{equation} \label{Berry-curvature}
\mathbf{\Omega} ({\bf k} )  =\frac{m  v_z   k_\rho/\Lambda}{[(\frac{2 k_0 }{\Lambda}q_\rho)^2+v_z^2k_z^2+ m^2]^{\frac{3}{2}}}\e_\phi  ,
\end{equation}
where we have neglected terms of higher order in $q_\rho$ and $k_z$.
Here,  $q_\rho=k_\rho-k_0$ is the radial distance from the Dirac ring.
The Berry curvature is peaked at $(q_{\rho} , k_z)=(0,0)$ and   points in opposite
directions at opposite sides of the Dirac ring (Fig.~\ref{Current}). The latter is a consequence
of  time-reversal symmetry, which requires that $\mathbf{\Omega} ( {\bf k} ) =  - \mathbf{\Omega} ( - {\bf k} )$.
 
From Eq.~\eqref{Berry-curvature} we  compute the transverse current contributed by states with
momentum angle $\phi$ by performing the momentum integral   over the  cylindrical coordinates $k_{\rho}$ and $k_z$.
 Assuming that the chemical potential $E_{\textrm{F}}=\mu$ lies within the conduction band, just above the gap opened by $m \sigma_1$,
 we obtain the   $\phi$-dependent Hall current  
 \begin{equation}
\mathbf{j}_{\textrm{t},\phi} \label{Current-density}
\simeq
\frac{e^2}{\hbar} \frac{ k_0}{8\pi^2}\left(1-\frac{m}{\mu}\right)\mathbf{E}\times\e_\phi, 
\end{equation}
where we have neglected terms of  order  $m^2$. 
For a derivation of Eqs.~\eqref{Berry-curvature} and~\eqref{Current-density}
we refer the reader to the supplemental material (SM)~\cite{Supp}. 
Interestingly, when the chemical potential $\mu$ is bigger than the gap energy $m$, the
transverse current $\mathbf{j}_{\textrm{t},\phi} $ is dominated by the first term, which  
originates from the parity anomaly.
Indeed, the first term of Eq.~\eqref{Current-density} is consistent with Eq.~\eqref{anomalous-current} as it differs
only by the  differential element $(k_0 / 2\pi) d\phi $ of the cylindrical coordinate system. 
Figure~\ref{Current} displays the distribution of the transverse currents $\mathbf{j}_{\textrm{t},\phi}$ (green arrows)
along the Dirac ring for a constant electric field applied
along the $\y$ and $\z$ directions. 
We observe that carriers on opposing sides of the Dirac ring flow into opposite directions transverse to the
electric field. This leads to an accumulations of charge on opposite surfaces of the DNLSM.

%%%%%%%%%%%%%%%%%%%%%%%%%%%%
\begin{figure}
\includegraphics[scale=0.40]{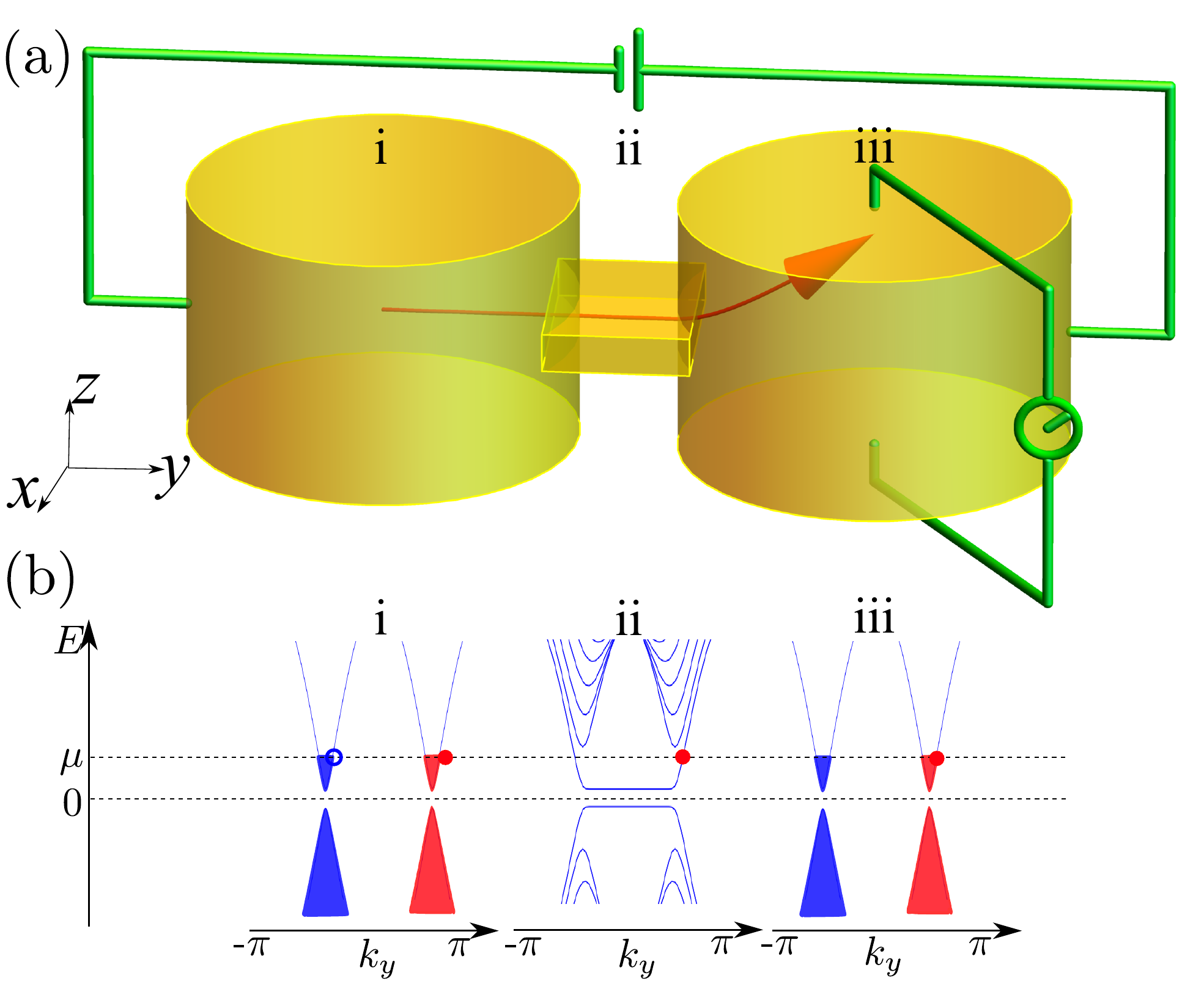} 
\caption{
(a)~The dumbbell device consists of two bulk regions (``i" and ```iii")  separated
by a constriction (``ii") with (001) surface states  (i.e., in-gap sates at the surface parallel to the $z$ direction). 
(b) Schematic dispersion relation for fixed $k_x=0$ in the bulk regions  and in the constriction.
An electron with $k_y >0$ (red filled circles) can be transmitted, while an electron with $k_y <0$ (open blue circles) is reflected. 
\label{Filter}
 }
\end{figure}
%%%%%%%%%%%%%%%%%%%%%%%%%%%%

\textit{Dumbbell filter device. } 
From the above analysis it is now clear that the parity anomaly in DNLSMs gives rise to transverse topological currents. 
However, since the currents contributed by modes on opposing sides of the Dirac ring have opposite sign, the total
transverse current vanishes (i.e., the anomaly cancels). Nevertheless, it is possible to detect anomalous currents by
use of a dumbbell filter device, which is based on a ballistic constriction with  (001) surface states [Fig.~\ref{Filter}(a)] (i.e., a constriction
in which the electronic states are confined along the $z$ direction).
To explain this, we consider a lattice version of the effective Hamiltonian~\eqref{model_ham}, which is given by
$\mathcal{H}_{\textrm{L}} = \sum_{\bf k} \Psi^{\dag}_{\bf k}  \mathcal{H}_{\textrm{L}}( {\bf k} ) \Psi_{\bf k}$, 
with $\Psi_{\bf k} = ( c_{\textrm{p}  {\bf k}} , c_{\textrm{d}  {\bf k}} )^{\textrm{T}}$ describing electrons 
in $p$ and $d$ orbitals, 
and~\cite{Nodal-Line-II} 
\begin{multline} \label{lattice_ham}
\mathcal{H}_{\textrm{L}} ( {\bf k} )=[ \mu_z-2t_\parallel(\cos k_x+\cos k_y)\\-2t_\perp \cos k_z]\sigma_3-2t^{\prime}_{\perp} \sin k_z \sigma_2+ m \sigma_1.
\end{multline}
Here, $\mu_z$ is an on-site energy, and $t_{\parallel}$, $t_{\perp}$, and $t^{\prime}_{\perp}$ represent intra- and inter-orbital 
hopping amplitudes on the cubic lattice.
The topologically nontrivial Berry bundle of $\mathcal{H}_{\textrm{L}}$ leads to the appearance of drumhead surface states.
This can be seen by deforming the green integration loop in Fig.~\ref{Nodalloop}(a) into two lines along the (001) axis (i.e., along the  $z$ direction), 
denoted by ``L$_i$"  in Fig.~\ref{Nodalloop}(b).
It follows from the bulk-boundary correspondence~\cite{vanderbilt_king_smith} that in-gap surface states appear at the (001) face   (i.e., at surfaces perpendicular to the $z$ direction) whenever $\nu[L_i] \ne 0$.
This corresponds to regions of the surface BZ  that are bounded by the projected Dirac ring,
since moving $L_i$  along transverse directions without crossing the Dirac ring preserves $\nu[L_i]$. 
% Since the drumhead surface states are of topological origin, their existence does not depend on the surface termination or any other microscopic details of the crystal surface.
%%

\begin{figure}
\includegraphics[scale=0.56]{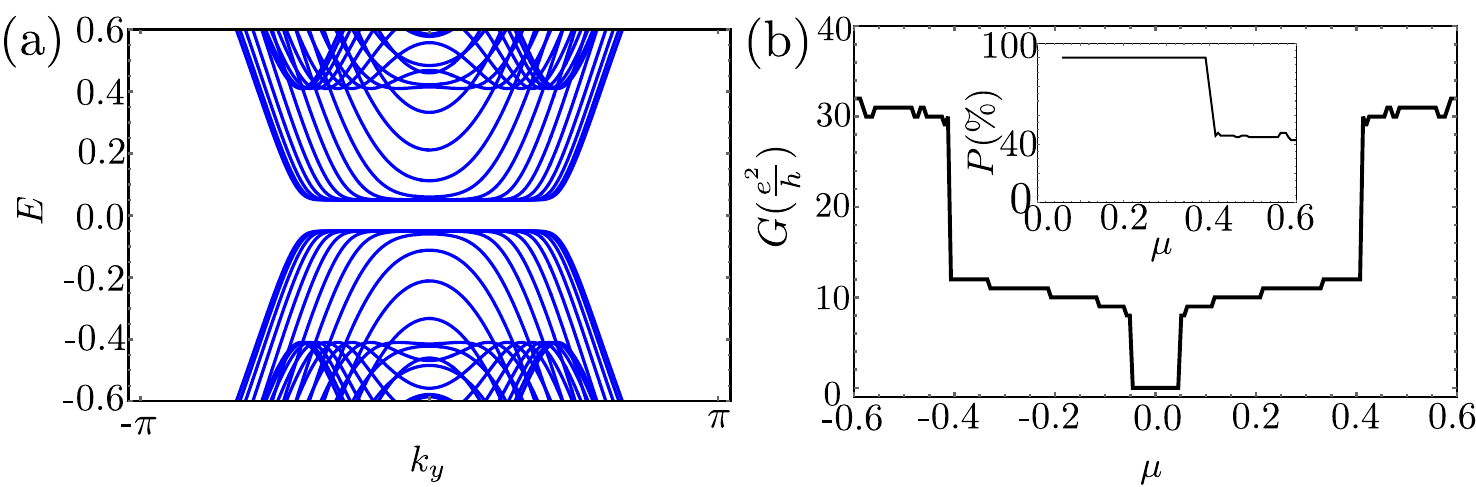} 
\caption{ \label{fig_conductance}
(a) Dispersion relation of the DNLSM $\mathcal{H}_{\textrm{L}}$, Eq.~\eqref{lattice_ham},  in   bar geometry with dimensions $N_x=20 $ and $N_z= 10$
and mass $m = 0.05$.
(b) Conductance $G$ for the dumbbell filter as a function of chemical potential $\mu$ in the constriction.
The inset shows the  polarization $P$.
 }
\end{figure}

The proposed dumbbell filter device  consists of two bulk regions connected by a ballistic constriction with drumhead surface states [Fig.~\ref{Filter}(a)].
All three parts of the dumbbell are made from the same nodal-line material, which should be relatively defect free. 
 Such a device could be manufactured, for example, using focused ion beam microfabrication~\cite{moll_nature_2016}.
The electronic states in the constriction are confined in the $z$ direction, such that their low-energy spectrum is dominated by the drumhead surface states.
We show the dispersion relation of the constriction with dimensions $N_x=20 $ and $N_z= 10$ in Fig.~\ref{fig_conductance}(a),  which
reveals that all states with energies within the  interval $-0.4 \lesssim E  \lesssim 0.4$ are surface states (cf.~Fig.~S3(a) in SM~\cite{Supp}).
When a voltage is applied across the device, a current   passes through the constriction, whose
conductance  is given by the  multi-channel Landauer formula~\cite{landauer_formula,kwant_package},  $G  = \frac{e^2}{h} \sum_{\mu, \nu} 
\left| t_{\mu \nu}   \right|^2$, where $t_{\mu \nu} $ are the transmission coefficients.
Assuming that the chemical potential $\mu$   lies slightly above the gap energy, 
transport through the constriction is mediated mainly by the modes of the drumhead surface states. 
Indeed, as shown in Fig.~\ref{fig_conductance}, for  $  | \mu  | \lesssim 0.4$ the current flows entirely within the surface states, leading 
to plateaus of quantized conductance with steps in multiples of $\frac{e^2}{h}$.
For $|\mu| \gtrsim 0.4$, however, bulk modes start to contribute.
Since the right propagating surface modes all have positive $k_y$, only electrons from the right half of the Dirac ring [red  in Fig.~\ref{Filter}(b)]
with $k_y >0$ can pass through the constriction. Electrons from the left half of the Dirac ring [blue in Fig.~\ref{Filter}(b)], however, are
reflected. Therefore, the dumbbell device acts as a filter for modes with $k_y >0$. The effectivness of the filter can be estimated by
the polarization $P =  G_{\textrm{surf}} / G_{\textrm{tot}}$, where   $G_{\textrm{tot}}$ and $G_{\textrm{surf}}$ denote the total conductance and
the conductance contributed by the surface modes, respectively. We find that  $P$ is close to 100\% for $| \mu  | \lesssim 0.4$, while it decreases
once bulk modes start to mix in [Fig.~\ref{fig_conductance}(b)]. For additional simulations see SM~\cite{Supp}. 

%%%%%%%%%%%%%%%%%%%%%%%%%%%%
%%%%%%%%%%%%%%%%%%%%%%%%%%%%

Now,  since the electric field is oriented along the $y$ direction in the dumbbell device,
electrons with $k_y >0$ give rise to a transverse current
 that flows upwards along the $z$ direction [Fig.~\ref{Current}(a)].
Thus,  a voltage difference develops between the upper and lower surfaces of the right weight plate of Fig.~\ref{Filter}(a).
This voltage difference can be measured experimentally and is 
a clear signature of the parity anomaly in DNLSMs.

\textit{Conclusion. } 
The anomaly-induced currents are robust to small perturbations, since they are of topological origin.
The same  applies to the dumbbell device, as its properties originate from topologically protected surface states. 
Hence, the topological currents are observable even in systems with small spin-orbit coupling, finite dispersion of the nodal ring, as well as moderately strong disorder, as discussed in the SM~\cite{Supp}.
Regarding experimental realizations of our proposal,  the 
hexagonal pnictides CaAgAs and CaAgP~\cite{okamoto_takenaka_JPSJ_16,emmanouilidou_PRB_17} are
particularly promising candidate materials, because they are available in single crystal from
and exhibit just a single Dirac ring at the Fermi energy~\cite{emmanouilidou_PRB_17,wang_CaAgAs_PRB_17,2017arXiv170807814N,2017arXiv170806874T}.  In the SM~\cite{Supp} we give some estimates for the optimal geometry of a dumbbell device made out of  CaAgP.

While the observation of the parity anomaly in DNLSMs would be of fundamental interest,
the dumbbell device used for this purpose could also lead to new electronic devices, such as a topological current rectifier. 
We anticipate that similar devices could also be realized in  Dirac or Weyl semi-metals, whose Fermi arc
surface states could be used as a valley filter.  
 
\begin{acknowledgements}
We thank Philip Moll and Ali Bangura for useful discussions. A.P.S. is grateful to the KITP at UC Santa Barbara for hospitality during the preparation of this work. This research was supported in part by the National Science Foundation under Grant No. NSF PHY-1125915.
 \end{acknowledgements}
\bibliographystyle{apsrev4-1}

\bibliography{NL-Trans-Ref}

\begin{thebibliography}{70}
\bibitem [{Sym()}]{SymClass}%
  \BibitemOpen
  \href@noop {} {}\bibinfo {note} {Hamiltonian~\eqref{model_ham} describing our
  nodal-line system belongs to symmetry class AI in the ten-fold
  classification. However, only in the presence of $PT$ symmetry, with
  $(PT)^2=+1$ are the nodal lines stable and topologically
  protected~\cite{Zhao-Schnyder-Wang-PT}.}\BibitemShut {Stop}%
\end{thebibliography}

 %%%%%%%%%%%%%%%%%%%%%%%%%%%%%%%%%%%%%%%%%%%%%%%%%%
%%%%%%%%%%%%%%%%%%%%%%%%%%%%%%%%%%%%%%%%%%%%%%%%%%

 \end{document}